# Turbulent formation of protogalaxies at the end of the plasma epoch: theory and observations

Rudolph E. Schild[1,2]

[1] Center for Astrophysics, 60 Garden Street, Cambridge, MA 02138, USA
[2] rschild@cfa.harvard.edu

and

Carl H. Gibson [3,4]

[3] University of California San Diego, La Jolla, CA 92093-0411, USA
[4] cgibson@ucsd.edu, http://sdcc3.ucsd.edu/~ir118

**ABSTRACT**

The standard model of gravitational structure formation is based on the Jeans 1902 acoustic theory, neglecting nonlinear instabilities controlled by viscosity, turbulence and diffusion.  A linear instability length scale equal to the sound speed times the gravitational freefall time emerges.  Because the Jeans scale $L_J$ for the hot primordial plasma is always slightly larger than the scale of causal connection $ct$, where $c$ is the speed of light and $t$ is the time after the big bang, it has been assumed that no plasma structures could form without guidance from a cold (so $L_{J\ CDM}$ is small) collisionless cold-dark-matter CDM fluid to give condensations that gravitationally collect the plasma.  Galaxies by this CDM model are therefore produced by gradual hierarchical-clustering of CDM halos to galaxy mass over billions of years, contrary to observations showing that well formed galaxies existed much earlier.  No observations exist of CDM halos.  However,  Gravitational instability is non-linear and absolute, controlled by viscous and turbulent forces or by diffusivity at Schwarz length scales smaller than $ct$.  Because the universe during the plasma epoch is rapidly expanding, the first structures formed were at density minima by fragmentation when the viscous-gravitational scale $L_{SV}$ first matched $ct$ at 30,000 years to produce protosupercluster voids and protosuperclusters.  Weak turbulence produced at expanding void boundaries guide the morphology of smaller fragments down to protogalaxy size just before transition to gas at 300,000 years.   The size of the protogalaxies reflect the plasma Kolmogorov scale with a linear and spiral morphology predicted by the Nomura direct numerical simulations and confirmed by deep field space telescope observations.  On transition to gas the kinematic viscosity decreases by ten trillion so the protogalaxies fragment as Jeans scale clouds, each with a trillion earth-mass planets, as predicted by Gibson 1996 and observed by Schild 1996.  The planets promptly form stars near the cores of the protogalaxies, but as the universe cools the planets gradually freeze to form the galaxy dark matter.  Space telescope observations of the most distant  galaxies confirm the Kolmogorov scale size and Nomura geometry of the protogalaxies.  High resolution images of planetary nebula and supernova remnants reveal thousands of frozen hydrogen-helium dark matter planets with large atmospheres evaporated by such events.  Galaxy mergers show frictional trails of young globular clusters formed in place, proving that dark matter halos of galaxies consist of dark matter planets in metastable clumps.  Galaxy evolution is guided by friction from this galaxy dark matter.

**INTRODUCTION**

The standard model of gravitational structure formation and the formation of galaxies is based on a linear acoustic instability theory proposed by James Hopwood Jeans in 1902 in a lengthy monograph titled "The stability of a spherical nebula" transmitted to the Philosophical Transactions of the Royal Society of London Series A (Vol. 1999, pp. 1-53) through G. H. Darwin, son of Charles

Darwin. The purpose of the work was to understand how the solar system might have formed from a gas cloud. Galaxies and the expansion of the universe were not known at that time. The alternative theory due to G. H. Darwin in his 1889 paper (Vol. 180, pp. 1-69) was titled "On the mechanical conditions of a swarm of meteorites, and on theories of cosmogony". Neither of the theories are correct as a description of how either the solar system or galaxies were formed. Darwin's theory focusing on meteorites [1] is a better guide than Jeans' in both cases since accretion within clumps of primordial planets and effective viscosity from planet collisions are crucial and the collapse of gas clouds following Jeans 1902 [2] is irrelevant. The Jeans acoustic scale has only one known application, which is to set the mass of globular star clusters at the plasma to gas transition (decoupling), but it is for different reasons than described by Jeans. Modern fluid mechanical concepts [3-19] applied to cosmology are termed hydro-gravitational-dynamics HGD. Relevant length scales are summarized in Table 1.

The fluid mechanics of Jeans 1902 is that of the 19$^{th}$ century. Jeans started with the Euler equations that neglect viscous forces, used linear perturbation stability analysis that neglects turbulence forces, and took no account of diffusivity effects arising from weakly collisional non-baryonic dark matter that probably constitutes about 97% of the rest mass of the universe where the universe is flat with density approximately equal to the present critical density of $\rho_{Crit} = 10^{-26}$ kg m$^{-3}$ with no dark energy or cosmological constant $\Lambda$. Jeans' assumptions lead to linear acoustic momentum conservation equations for the gas, so the Jeans 1902 acoustic instability scale is $L_J = V_S / (\rho G)^{1/2}$, where $V_S$ is the speed of sound, $\tau_g = (\rho G)^{-1/2}$ is the gravitational free fall time, $\rho$ is the density and G is Newton's constant of gravitation.

The Jeans criterion for gravitational structure formation is that a large cloud of gas with density $\rho$ is unstable to gravitational structure formation only on scales larger than $J_H$. By this criterion it is impossible to make any structures at all in the plasma epoch of the universe because the sound speed $V_{S-plasma} = c / 3^{1/2}$ is so large that $L_J > L_H$, where $L_H = ct$ is the horizon scale or scale of causal connection at time $t$ after the big bang and $c$ is the speed of light. Even after the plasma turns to gas at $t \approx 10^{13}$ seconds (300,000 years) the Jeans mass is an enormous million solar masses so the first stars cannot form until much later and planets can never form from gas at average temperatures because the density of the expanding universe decreases more rapidly than the sound speed with time. The problem with the Jeans criterion is that it is simply wrong [10]. Gravitational instability is absolute not linear [11], meaning that in a fluid with density fluctuations, structure will immediately begin to form at all scales of available density fluctuations unless prevented by the molecular diffusivity of the nearly collisionless non-baryonic dark matter or by viscous or turbulence forces. Mass moves toward density maxima and away from density minima at all scales unless prevented by fluid or other forces or by diffusivity.

When the plasma turns to gas it becomes a fog of planets [10] that become galaxy dark matter, as observed [20]. The only known form of non-baryonic dark matter NBDM is neutrinos, but there may be other neutrino-like particles with neutrino-like mass ~ 10$^{-35}$ kg and neutrino-like collision cross section ~ 10$^{-37}$ m$^2$ [11] formed in the early universe at temperatures unavailable in laboratories or there may be other flavors of neutrinos than known from present estimates. The evidence indicates the existence of something non-baryonic and quite massive that prevents the disintegration of galaxy clusters by gravitational forces, just as baryonic dark matter is needed to inhibit the disintegration of individual galaxies by centrifugal forces, but at this time no one knows what it is.

Density perturbations exist in the plasma because the big bang was triggered by a form of turbulent combustion at Planck temperature 10$^{35}$ K termed the Planck-Kerr instability [12,13]. Planck particles and anti-particles appear by quantum tunneling and form the Planck scale equivalent of posi-tronium. Positronium is produced by supernova temperatures of $10^{10}$ K sufficient to cause electron positron pair production, where the antiparticle pairs form a relatively stable combinations in orbit. Prograde captures give 42% release of the Planck particle rest mass energy which can only go into producing more Planck particles to fuel similar interactions. This gives a turbulent big bang with Reynolds number of about a million, terminated when cooling permits formation of quarks and gluons and a strong increase of viscosity and viscous stress that can cause an exponential increase in space to approximately a meter size in $10^{-33}$ seconds from $10^{-27}$ meters ($10^8$ Planck lengths). The speed is about $10^{25}$ $c$. This inflation epoch produces the first fossil turbulence because all the turbulent temperature fluctuations of big bang turbulence are stretched beyond the scale of causal connection $L_H = ct$.

These fossils of big bang turbulence then trigger the formation of density fluctuations in the hydrogen and helium formed in the first three minutes by nucleosynthesis. The density fluctuations then seed the first gravitational structure formation [14]. The mechanism is not by the Jeans criterion but under the control of viscosity, turbulence or particle diffusivity, depending on which of three Schwarz scales is largest for the fluid in question. Because the collision length for non-baryonic dark matter is larger than $L_H = ct$ for the plasma epoch, only viscous forces or turbulence forces can prevent structure formation. Fragmentation of the non-baryonic dark matter occurs after the end of the plasma epoch, so this material is nearly irrelevant to the first structure formation, which begins when the horizon scale matches the Schwarz viscous scale at about $10^{12}$ seconds after the big bang (30,000 years) and the mass scale is that of super-clusters. The plasma epoch is dominated by viscous forces with only weak turbulence because the photon viscosity is very large. Photons scatter from the free electrons of the plasma and transmit momentum because the electrons strongly couple to the ions to maintain electrical neutrality. Reynolds numbers are close to critical values so the Schwarz viscous and turbulent scales are nearly identical. The first structures to form are voids because the universe is expanding. Condensations occur only after decoupling. The voids expand at the sonic speed of the plasma epoch which is $c/3^{1/2}$, so we have significantly larger super-voids than super-clusters as discussed in the following sections [15].

Because observations showed early structure must have occurred, cosmologists resorted to a *deus ex machina* solution (magic): they invented cold dark matter in order to create the present standard cosmological model. Following Jeans, it was assumed that a cold non-baryonic material must exist with sound speed sufficiently small that gravitational condensations in the plasma epoch could occur; that is, with $L_{J_{CDM}} < L_H$. Seeds of CDM were assumed to condense to provide gravitational potential wells into which the baryonic plasma would fall. The seeds would magically merge and stick over time, collecting more and more of the baryonic material until stars could form, and the clumps could cluster to larger and larger scales to produce galaxies and galaxy clusters and last superclusters of galaxies. This is hierarchical clustering, so the standard model is often called CDMHC. No CDM halo has ever been observed. In the following we suggest they do not exist and that CDM does not exist. Neither does the latest addition to the standard cosmological model; that is, dark energy and the cosmological constant $\Lambda$ [16–19].

Instead, the plasma fragmentation beginning at $10^{12}$ s continues to smaller and smaller scales. Galaxies represent the smallest mass gravitational structures formed in the plasma before decoupling at $10^{13}$ s (300,000 years). Galaxy morphology reflects the turbulence existing in the plasma when they were formed. Proto-galaxies fragmented along turbulent vortex lines at the Kolmogorov scale of the plasma and with a chain and spiral morphology demonstrated by the direct numerical simulations of Nomura. This morphology and a consistent proto-galaxy length scale termed the Nomura scale is observed in a variety of Hubble Space Telescope observations discussed in the rest of this chapter [15].

At decoupling the kinematic viscosity $\nu$ decreased from photon viscosity values of ~ $10^{26}$ m$^2$ s$^{-1}$ to hot gas values of $10^{13}$ m$^2$ s$^{-1}$, a factor of ten trillion. This decreased the mass scale of fragmentation from that of proto-galaxies to that of planets. Simultaneously the gas proto-galaxies fragmented at the Jeans mass of globular star clusters, giving hot gas clouds of planet mass in million solar mass clumps. As the universe expands it cools. The gas clouds of hydrogen and helium eventually begin to freeze to form the galaxy dark matter, which is proto-globular-star-cluster PGC clumps of primordial-fog-particle PFP planets [10] as observed [20]. All stars form in PGCs from the trillion planets contained by a binary accretion process that gives larger and larger Jovian planets. The first stars to form were therefore small and long-lived, as observed in ancient globular star clusters. Structure formation from fluid mechanics naturally explains globular star clusters, which have always been a mystery. In standard ΛCDMHC cosmology, the first stars to form are superstars that re-ionize the universe to explain why so much hydrogen is missing in high red-shift quasar spectra. In our hydro-gravitational-dynamics HGD cosmology the missing hydrogen is tied up in frozen planets, the superstars (Population III) never happened, and Re-ionization of the Universe never happened [15].

Table 1. Length scales of gravitational instability

| Length Scale Name | Definition | Physical Significance |
|---|---|---|
| Jeans Acoustic | $L_J = V_S/(\rho G)^{1/2}$ | Acoustic time matches free fall time |
| Schwarz Viscous | $L_{SV} = (\gamma\nu/\rho G)^{1/2}$ | Viscous forces match gravitational forces |
| Schwarz Turbulent | $L_{ST} = (\varepsilon/[\rho G]^{3/2})^{1/2}$ | Turbulent forces match gravitational forces |
| Schwarz Diffusive | $L_{SD} = (D^2/\rho G)^{1/4}$ | Diffusive speed matches free fall speed |
| Horizon, causal connection | $L_H = ct$ | Range of possible gravitational interaction |
| Nomura, smallest plasma piece | $L_N = 10^{20}$ m | Proto-galaxy-fragmentation on vortex lines |

$V_S$ is sound speed, $\rho$ is density, G is Newton's constant, $\gamma$ is the rate of strain, $\nu$ is the kinematic viscosity, $\varepsilon$ is the viscous dissipation rate, D is the diffusivity, $c$ is light speed, $t$ is time.

## THEORY

Figure 1 shows the sequence of gravitational structure formation events leading to the present time according to hydro-gravitational-dynamics HGD cosmology.

The big bang mechanism is one of turbulent combustion limited by gluon viscosity, where the maximum Taylor microscale Reynolds number Re$_\lambda$ = 1000 occurred at t = $10^{-33}$ seconds at $10^8$ L$_P$, where L$_P$ = $10^{-35}$ is the Planck scale of quantum-gravitational instability [12,13]. The only particles possible during this epoch of high temperatures were Planck particles and Planck anti-particles that interact in the manner of electron-positron pair production to efficiently produce more Planck particles until the event cools enough for quarks and gluons to appear. Large negative Reynolds stresses expand space according to Einstein's theory of general relativity. During the big bang turbulence epoch, indicated in Fig. 1a by the red star, the kinematic viscosity $\nu$ is the mean free path for particle collisions L$_P$ times the speed of the particle. Even though the particle speeds were light speeds $c$ at $t = 0$, the collision distances were small, giving a small initial kinematic viscosity $\nu_0 = 3\times10^{-27}$ m$^2$ s$^{-1}$. The Planck scale Reynolds number is near critical so L$_P$ matches the Kolmogorov and Batchelor scales [3] of big bang turbulence and turbulent mixing.

The end of the big bang turbulence epoch occurred when the temperatures and particle speeds decreased as the Reynolds numbers increased till quarks and gluons could appear. This phase change is termed the strong force freeze out, with length scale L$_{SF}$ = $10^8$ L$_P$, or $10^{-27}$ m. Gluons are particles that carry momentum for the quarks, similar to photons carrying forces from momentum differ-ences between the electrons and protons of the plasma epoch. It is assumed that the gluon viscosity is dramatically larger than the Planck viscosity, just as the photon viscosity is dramatically larger than the proton-proton collision viscosity of the plasma epoch, so that large negative gluon viscous stresses will accelerate the expansion of space exponentially during the inflation epoch indicated by the blue triangle. Other mechanisms such as the Guth false vacuum may be important for driving inflation. Whichever mechanism is dominant, many observations support an inflationary event.

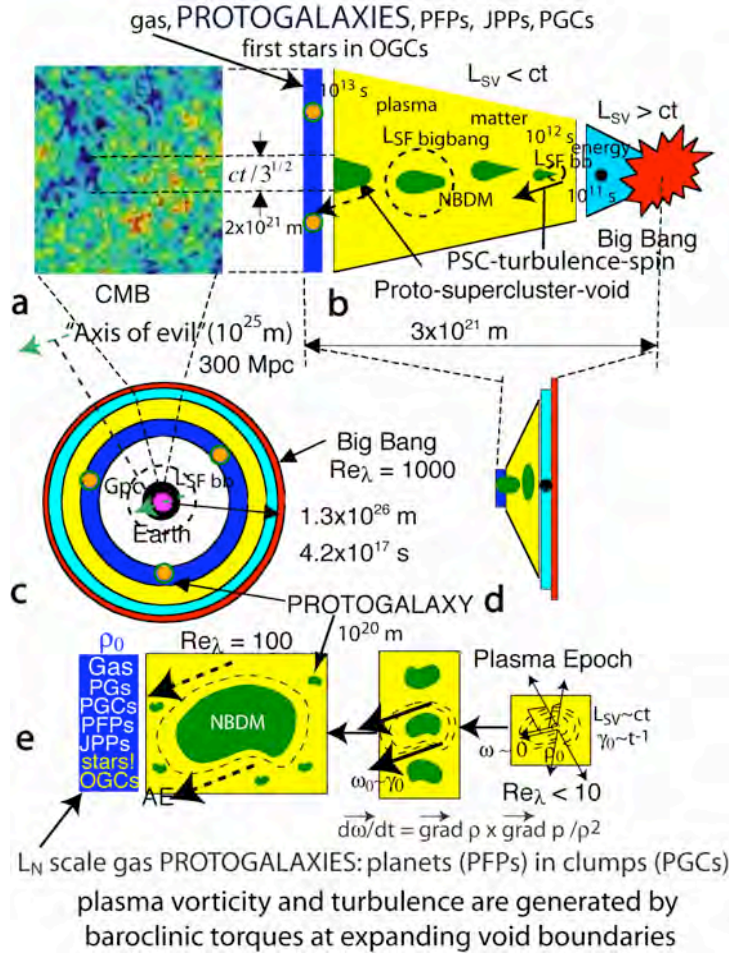

Figure 1. Protogalaxy formation at the end of the plasma epoch by hydro-gravitational-dynamics HGD theory. a. Cosmic Microwave Background temperature anisotropies reflect structures formed in the plasma epoch. b. From HGD the photon viscosity of the plasma epoch prevents turbulence until the viscous Schwarz scale $L_{SV}$ becomes less than the Hubble scale (horizon scale, scale of causal connection) $L_H = ct$, where $c$ is the speed of light and $t$ is the time. The first plasma structures were proto-super-cluster voids and proto-super-clusters at $10^{12}$ seconds (30,000 years). c. Looking back in space is looking back in time. Proto-galaxies were the last fragmentations of the plasma (orange circles with green halos) at $10^{13}$ seconds. d. The scale of the gravitational structure epoch is only $3 \times 10^{21}$ m compared to present super-cluster sizes of $10^{24}$ m and the largest observed super-void scales of $10^{25}$ m. e. Turbulence in the plasma epoch is generated by baroclinic torques on the boundaries of the expanding super-voids.

During inflation, turbulent temperature fluctuations are fossilized by stretching beyond the scale of causal connection $L_H$. The mass-energy and entropy vastly increase, presumably by the rapid stretching of superstrings with Planck tension $c^4/G = 1.2 \times 10^{44}$ kg m s$^{-2}$ [21]. The black dot in the inflation epoch symbolizes fossilized turbulent temperature fluctuations that can guide nucleosynthesis and preserve big bang turbulence information for testing using CMB temperature anisotropy patterns. In a period of $10^{-27}$ seconds the size of the big bang universe increases from $10^{-27}$ m to a few meters with density exceeding $10^{80}$ kg m$^{-3}$. The mass-energy of the universe within our present horizon scale $L_H = 10^{26}$ m is about $10^{52}$ kg, about $10^{-40}$ of the mass-energy created by the big bang. The mass-energy of the universe before inflation was less than $10^{16}$ kg so the increase was by a factor of $10^{74}$. Guth describes this as the ultimate free lunch [21].

The transition between energy domination and mass domination occurred at about $10^{11}$ s, the beginning of the plasma epoch shown in Fig. 1b. The plasma is absolutely unstable and will fragment once the Schwarz viscous scale $L_{SV}$ matches the horizon scale $L_H$. This occurs at time $t = 10^{12}$ s when the density of a flat universe is $10^{-15}$ kg m$^{-3}$. Most of this density, about 97%, is non-baryonic dark matter, NBDM, presumably some combination of neutrinos. The rest is plasma. None of it is dark energy. This gives a primordial density $\rho_0$ of $4\times10^{-17}$ kg m$^{-3}$, which matches the density of old globular star clusters OGC. Because the universe continues to expand, fragmentation of the plasma at density minima is favored over condensation at density maxima. Voids appear in the plasma at $10^{12}$ seconds and proceed to expand as rarefaction waves at speeds limited by the speed of sound $c/3^{1/2}$ for $10^{13}$ seconds until decoupling. The material between the proto-super-cluster voids is proto-super-clusters with mass $10^{46}$ kg, that of a thousand galaxies. The proto-super-clusters never collapse gravitationally but continue to expand with the universe. This is a key point; instead of a messy chaotic formation scenario of mergers the expansion was tranquil and preserved primordial alignments and densities. The present scale of super-clusters is observed to be $10^{24}$ m.

Smaller scale fragmentations occur in the plasma epoch at density minima up to the time of plasma to gas transition (decoupling). The last fragmentation is to produce proto-galaxies at $10^{13}$ seconds, as shown in Fig. 1b. The morphology of the proto-galaxies is determined by weak turbulence generated by baroclinic torques on the surfaces of the super-voids as they expand. The direction of the spins is determined by the fossil strong force freeze-out density gradient at length scale $L_{SF}$. This spin direction determines the small wave-number spherical harmonic directions for the CMB, the spins of galaxies in the local group, and even the spin of the Milky Way. All are directed toward the "axis of evil" [17], which has length scales in quasar polarizations to a Gpc, or $3\times10^{25}$ m, as shown in Fig. 1c.

Figure 2 illustrates erroneous concepts of cold dark matter hierarchical clustering (CDMHC) theory, which is at present the standard cosmological model along with the equally erroneous concept of dark energy and a cosmological constant $\Lambda$ ($\Lambda$CDMHC). The standard model has no turbulence in the big bang but predicts small Gaussian scale-independent fluctuations and a white noise spectrum of CMB fluctuations from the big bang and inflation epochs. CDM attributes all structure formation in the plasma epoch to condensation and clustering of non-baryonic dark matter NBDM.

According to CDMHC seeds or halos of CDM form because their Jeans scale $L_J$ is less than $L_H$. This is physically impossible because the diffusivity of any form of NBDM will cause the diffusive Schwarz scale $L_{SD}$ of the material to exceed $L_H$. Even if a seed could form it could not stick to another seed because sticking requires collisions of the CDM particles, no matter how cold they might be. In Fig. 2 each erroneous CDM concept is indicated by a red X.

It is claimed by CDMHC that the baryonic matter plasma is inviscid and falls into the CDM halos where it oscillates without viscous damping to give the acoustic peaks observed in the CMB temperature anisotropy spectrum. However, the plasma is far from inviscid, with a photon viscosity of $\nu = 4\times10^{26}$ m$^2$ s$^{-1}$ [11]. Even if CDM halos existed, the plasma is far too viscous to enter the halo and oscillate acoustically.

Other erroneous concepts of CDMHC are listed in Fig. 2. The CDM halos gradually cluster and collect gas for 300 My of dark ages ($10^{16}$ s) in mini-galaxies which form the first superstars, one per mini-galaxy, in an enormous set of supernovae that re-ionizes all the gas formed at decoupling. The re-ionization concept is unnecessary to explain the lack of UV-absorbing gas as observed in quasar spectra because the gas is sequestered as dark matter planets in PGC clumps.

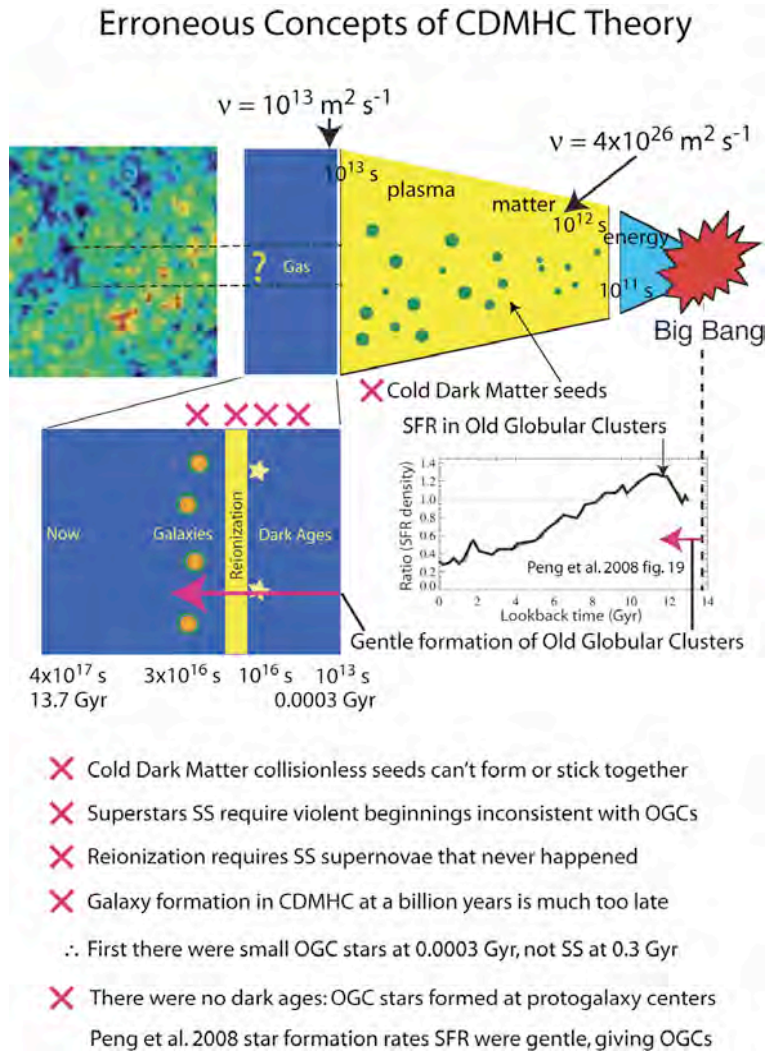

Figure 2.  Galaxy formation by the standard ΛCDM model is impossible to reconcile with observations and fluid mechanical theory (see text).  Failed aspects of the model are indicated by red Xs.  Numerical simulations invent "Plummer forces" to simulate CDM seed sticking.

The first CDMHC galaxies to form occur much too late to be consistent with observations, at about a billion years after the big bang ($3 \times 10^{16}$ s).  The star formation rate SFR in Peng et al. 2008 Fig. 19 [23] cannot be superstars because superstars can only occur supported by violent turbulence stresses that would prevent formation of the very small stars observed to exist in old globular star clusters OGC.

From HGD and the mass of small stars it is easy to show from the Schwarz length scale criteria of Table 1 that the maximum viscous dissipation rate ε to permit the formation of a small OGC star is $10^{-13}$ m$^2$ s$^{-3}$.  However, strong turbulence is required with ε values of $5 \times 10^{-4}$ m$^2$ s$^{-3}$ to permit formation of superstars and the re-ionization of the universe.  If such large ε values existed before the first star formed, no old-globular-clusters OGCs could have formed because their small stars would be inhibited by turbulence produced as NBDM CDM halos filled with baryonic gas.  Since OGCs are observed in all galaxies, the superstars and re-ionization concepts of CDMHC must be ruled out.

# OBSERVATIONS

The fifth year WMAP observations of the CMB have been released, as shown in Figure 3. Fig. 3a shows the Dunkley et al. 2008 5$^{th}$ year WMAP spherical harmonic temperature anisotropy data, with un-binned samples as light dots in the background [24].

The range of the un-binned samples in Fig. 3a is a measure of the intermittency (non-Gaussianity) of the turbulence associated with the samples. The range indicated by circles and arrows is smaller at the sonic peak than at higher wavenumbers. This is attributed to the weak turbulence produced in the plasma at the boundaries of the expanding supervoids. The wider scatter at high wavenumbers k is attributed to the greater intermittency of big bang turbulence, which has a higher Reynolds number. An Obukhov-Corrsin turbulent mixing spectrum $\beta\chi\varepsilon^{-1/3}k^{-5/3}$ is fitted to the lowest wavenumber CMB spectrum, where $\beta$ is a universal constant, $\chi$ is the dissipation rate of temperature variance and $\varepsilon$ is the dissipation rate of velocity variance. It is assumed that the measured spherical harmonic CMB spectrum multiplied by $\ell(1+\ell)$ is not much different than a turbulent dissipation spectrum $k^2\phi$.

The red solid curve labeled CDM+$\nu$=0 in Fig. 3a is a CDM model without viscosity forced by *ad hoc* numerical assumptions to fit the data. Photon viscosity would not permit the baryonic plasma to enter the CDM potential wells even if they were physically possible, and would not permit any sonic oscillations because of viscous damping. Therefore the Obukhov-Corrsin green-dashed dissipation spectrum from big bang turbulence theory labeled CDM+$\nu$ should apply.

Evidence of turbulence from the expansion of super-voids producing baroclinic torques and vorticity at void boundaries is shown by a series of turbulence statistical parameters computed by Bershadskii and Sreenivasan from the low wavenumber CMB spectrum in Fig. 3b, 3c, and 3d and compared to the same parameters computed for atmospheric, laboratory and numerical simulations of turbulence over a wide range of Reynolds numbers.

Figure 3b compares Weiner-filtered CMB data with extended self similarity turbulence data for various order structure functions. The agreement clearly demonstrates the CMB temperature anisotropies in the indicated angular size range 0.5 to 4 degrees studied were produced by turbulence. From HGD the turbulence reflects baroclinic torques at gravitationally driven expanding void boundaries.

Figure 3c compares the same CMB and turbulence data using a statistical parameter to test for Gaussianity. Again, the agreement with turbulence is remarkable. Both sets of CMB and turbulence data deviate from Gaussianity in the same way in the angular size range studied.

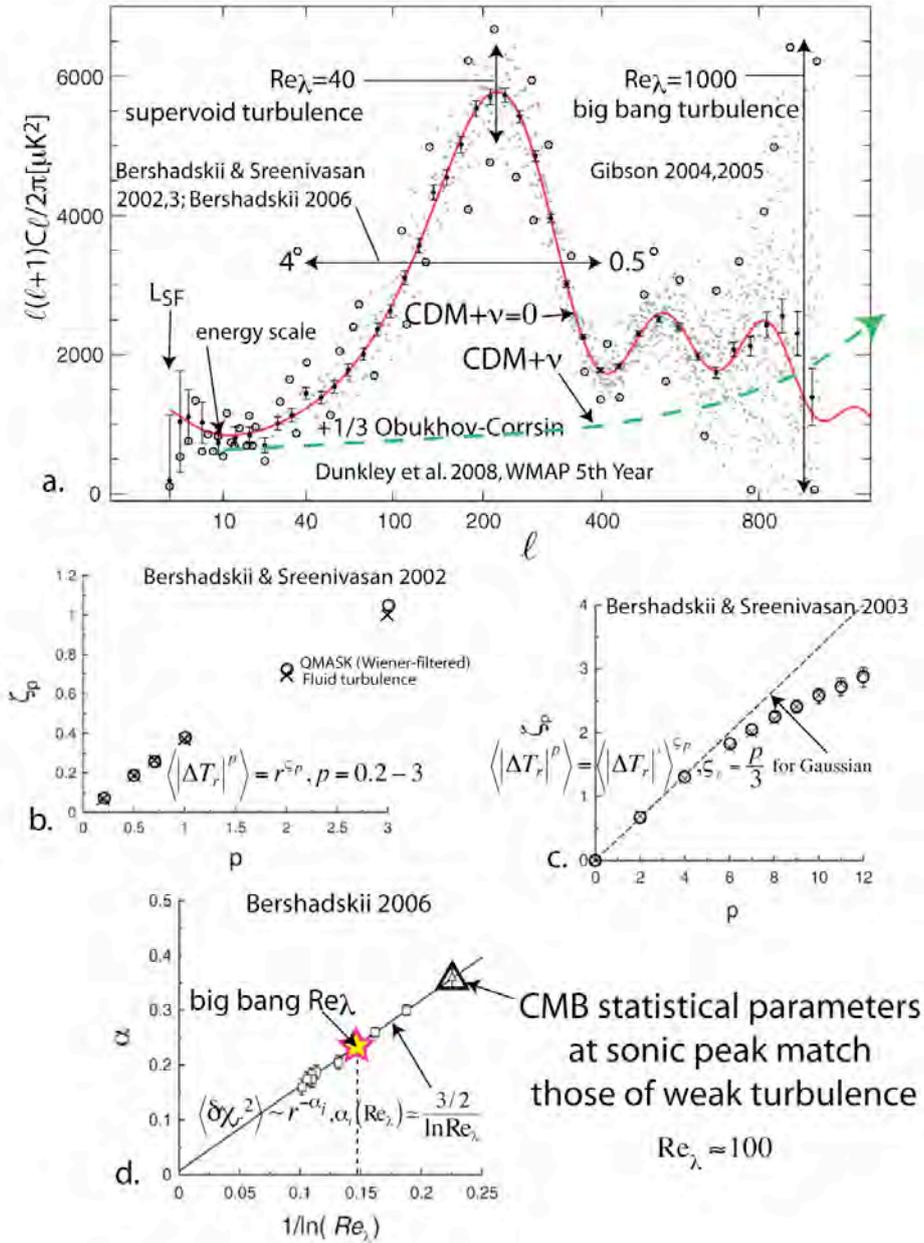

Figure 3. a. Spherical harmonic spectrum of CMB temperature anisotropies from the fifth year WMAP observations, Dunkley et al. 2008 [24]. The large sonic peak at wavenumbers near 200 reflects weak turbulence produced by gravitationally driven protosupervoid expansion according to Fig. 1, not inviscid CDM (solid red). The dashed green line represents a viscous CDM model; that is, with no sonic signal and should match the +1/3 Obukhov-Corrsin turbulent mixing dissipation spectrum of big bang turbulence. Turbulence parameters are compared to CMB statistics in b. [25], c. [26] and d. [27] as described in the text.

Figure 3d shows an estimate of the sonic peak Reynolds number estimated by Bershadskii 2006 to be about Re$\lambda$ ~ 100. A more precise estimate is Re$\lambda$ = 40 as shown near the peak, using the expression derived for the temperature anisotropy variance as a function of separation distance r between sampling points. The power law exponent is a function of Re$\lambda$. The value of Re$\lambda$ for the CMB is extracted using a series expansion of a(Re$\lambda$) for high Reynolds number as a function of 1/ln Re$\lambda$ . It turns out for turbulence that only the first term is important, as shown, so that the CMB data near the sonic peak gives Re$\lambda$ = 40 as indicated by the bold triangle. For comparison, the big bang turbulence Re$\lambda$ value of 1000 is shown by a bold star. There can be little doubt from the combination of evidence shown in Fig. 3 abcd that the plasma epoch was weakly turbulent in the range of scales including the sonic peak. As noted by Alexander Bershadskii (2001 personal communication to CHG) "the fingerprints of Kolmogorov are all over the sky".

As shown in Fig. 1e, the last stage of plasma fragmentations guided by weak turbulence is the formation of proto-galaxy-mass objects with a linear morphology reflecting vortex tubes of the turbulence where the rates-of-strain that enhance fragmentation of the plasma to form voids is maximum.

Figure 4 (bottom) is an image from the Hubble Space Telescope Advanced Camera for Surveys (HST-ACS) showing two proto-galaxy-chain clusters in the dim background of the Tadpole galaxy merger. Virtually all (94%) very dim galaxies from HST-ACS; that is, with magnitudes greater than 24, have such a linear morphology as expected from HGD [15]. The objects consist of bright nuclei separated by a dimmer glow that we interpret as proto-galaxies with a concentration of first-stars near the cores, with star-forming galaxy dark matter (planets in clumps) in between.

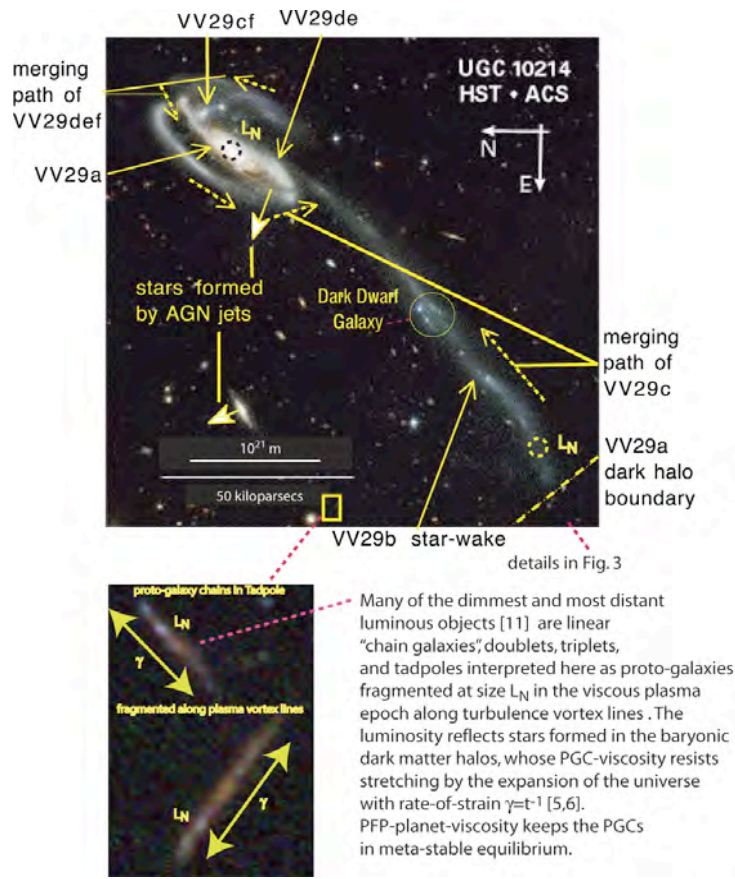

Figure 4. Tadpole galaxy merger (top), with background chains of protogalaxies shown in enlargement of small area at bottom center (bottom).

The star-clumps have a uniform size set by the turbulence and fluid mechanics to be the Nomura scale $10^{20}$ m (Table 1) with the Nomura morphology of weak turbulence [15].

Fig. 4 (top) shows the formation of star trails and dust trails in the baryonic dark matter halo of the Tadpole central galaxy VV29a as galactic objects VV29cdef merge in frictional spirals on the VV29a disk plane [19]. A sharp VV29a dark halo boundary is clear in the high-resolution HST-ACS images. The eponymous VV29b spiral filament shows numerous young globular star clusters [28] aligned in a direction pointing precisely at the point of frictional merger as the objects move through the dark matter halo triggering star formation by tidal forces in a $L_N$ scale diameter star wake. The size of VV29a dark matter halo diffused from the $L_N$ scale central core of old small stars is 8x10$^{21}$ m, or 0.9 Mpc.  From this evidence, the concept of frictionless tidal tails in galaxy mergers appears to be obsolete [29].

Figure 5 shows Seyfert galaxy NGC 6212 with its many closely associated quasi-stellar-objects QSOs, from Burbidge 2003 [30].  The radio-loud 3C 345 quasar is only 4.7 arcmin away, and has a companion QSO with the identical redshift 0.594.

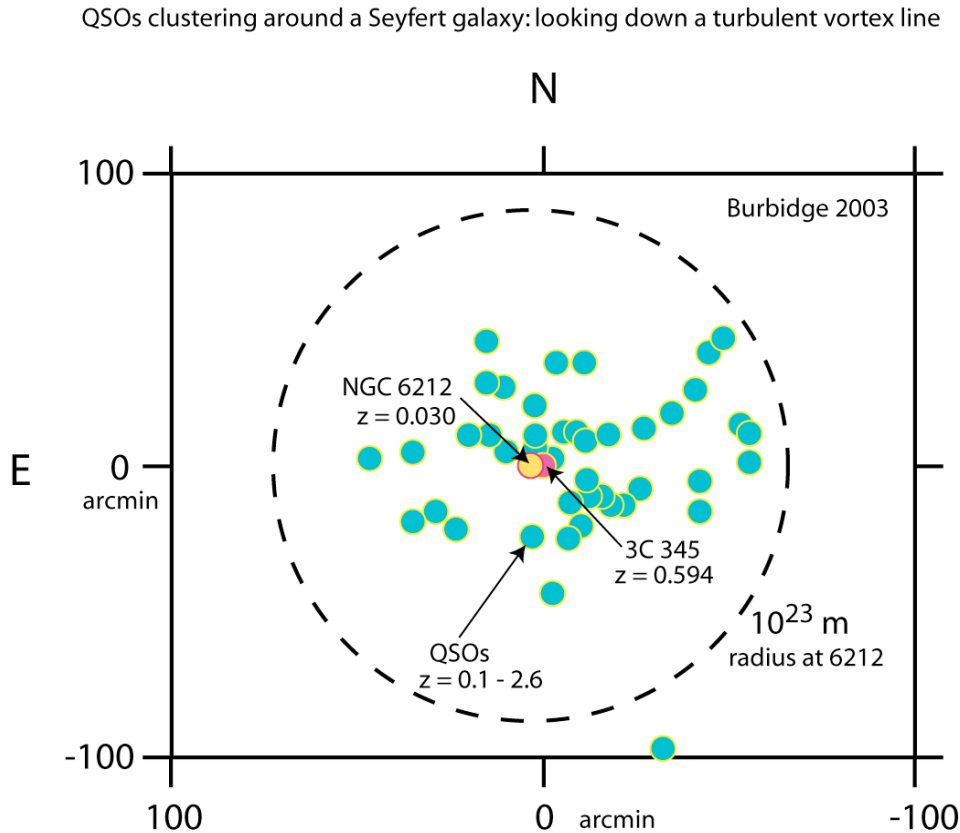

Figure 5.  QSOs with large redshifts are often observed clustering about active galaxies [30].  It has been strongly suggested that the QSOs have been emitted by the AGN with intrinsic redshifts and are actually very nearby.  An alternative from HGD is that the QSO galaxies and the line of sight are aligned with a cluster turbulent vortex line (see Fig. 1e).

Burbidge lists NGC 4258, NGC 2639, NGC 3516, NGC 1068, Arp 220 and NGC 3628 as additional examples of this phenomenon.  This can be understood from HGD as an example of galaxies and their spins determined during the plasma epoch by turbulence at the boundaries of expanding cluster voids.  If the line of sight coincides with the turbulent vortex line of the plasma epoch such alignments should occur.  No physical mechanism is known to permit emission of galaxies by AGNs with intrinsic redshifts.

**CONCLUSION**

A fluid mechanical analysis of gravitational structure formation termed hydro-gravitational-dynamics HGD shows that the CDMHC model is inconsistent with theory and observations and must be abandoned.  From HGD, gravitational structure formation is driven entirely by the baryonic matter of the plasma epoch.  The first structures are proto-super-cluster-voids and proto-super-clusters, followed by proto-galaxies at decoupling.  The proto-galaxies are formed in very weakly turbulent plasma, but possess the morphology of weak turbulence in that the protogalaxies are stretched along the vortex lines of the plasma turbulence with a diameter reflecting their viscous-inertial-vortex-gravitational origin at the Nomura scale 10$^{20}$ m.  Although our description of HGD seems to emphasize turbulence processes, the universe described is in fact more tranquil than previously envisioned, and can preserve primordial alignments seen in many cosmological structures, in particular proto-galaxies.

# REFERENCES

1. Darwin, G. H. 1889. On the mechanical conditions of a swarm of meteorites, and on theories of cosmogony, Phil. Trans., 180, 1-69.

2. Jeans, J. H. 1902. The stability of spherical nebula, Phil. Trans., 199A, 0-49.

3. Gibson, C.H. (1991). Kolmogorov similarity hypotheses for scalar fields: sampling intermittent turbulent mixing in the ocean and galaxy, Proc. Roy. Soc. Lond. A, 434, 149-164.

4. Gibson, C. H. (2006). Turbulence, update of article in Encyclopedia of Physics, R. G. Lerner and G. L. Trigg, Eds., Addison-Wesley Publishing Co., Inc., pp.1310-1314.

5. Gibson, C. H. (1981). Buoyancy effects in turbulent mixing: Sampling turbulence in the stratified ocean, AIAA J., 19, 1394.

6. Gibson, C. H. (1968a). Fine structure of scalar fields mixed by turbulence: I. Zero-gradient points and minimal gradient surfaces, Phys. Fluids, 11: 11, 2305-2315.

7. Gibson, C. H. (1968b). Fine structure of scalar fields mixed by turbulence: II. Spectral theory, Phys. Fluids, 11: 11, 2316-2327.

8. Gibson, C. H. (1986). Internal waves, fossil turbulence, and composite ocean microstructure spectra," J. Fluid Mech. 168, 89-117.

9. Gibson, C. H. (1999). Fossil turbulence revisited, J. of Mar. Syst., 21(1-4), 147-167, astro-ph/9904237

10. Gibson, C.H. (1996). Turbulence in the ocean, atmosphere, galaxy and universe, Appl. Mech. Rev., 49, no. 5, 299–315.

11. Gibson, C.H. (2000). Turbulent mixing, diffusion and gravity in the formation of cosmological structures: The fluid mechanics of dark matter, J. Fluids Eng., 122, 830–835.

12. Gibson, C.H. (2004). The first turbulence and the first fossil turbulence, Flow, Turbulence and Combustion, 72, 161–179.

13. Gibson, C.H. (2005). The first turbulent combustion, Combust. Sci. and Tech., 177: 1049–1071, arXiv:astro-ph/0501416.

14. Gibson, C.H. (2006). The fluid mechanics of gravitational structure formation, astro-ph/0610628.

15. Gibson, C.H. (2008). Cold dark matter cosmology conflicts with fluid mechanics and observations, J. Applied Fluid Mech., Vol. 1, No. 2, pp 1-8, 2008, arXiv:astro-ph/0606073.

16. Gibson, C.H. & Schild, R.E. (2007). Interpretation of the Helix Planetary Nebula using Hydro-Gravitational-Dynamics: Planets and Dark Energy, arXiv:astro-ph/0701474.

17. Schild, R.E & Gibson, C.H. (2008). Lessons from the Axis of Evil, axXiv[astro-ph]:0802.3229v2.

18. Gibson, C.H. & Schild, R.E. (2007). Interpretation of the Stephan Quintet Galaxy Cluster using Hydro-Gravitational-Dynamics: Viscosity and Fragmentation, arXiv[astro-ph]:0710.5449.

19. Gibson, C.H. & Schild, R.E. (2002). Interpretation of the Tadpole VV29 Merging Galaxy System using Hydro-Gravitational Theory, arXiv:astro-ph/0210583.

20. Schild, R. 1996. Microlensing variability of the gravitationally lensed quasar Q0957+561 A,B, ApJ, 464, 125.

21. Greene, B. 1999. The Elegant Universe, Superstrings, Hidden Dimensions, and the Quest for the Ultimate Theory, W. W. Norton & Compant, New York.

22. Guth, A. H. 1997. The Inflationary Universe, The Quest for a New Theory of Cosmic Origins, Helix Books, Addison-Wesley Pub. Co., Inc., New York.

23. Peng, E. W. et al. 2008, The ACS Virgo Cluster Survey. XV. The Formation Efficiencies of Globular Clusters in Early-Type Galaxies: The Effects of Mass and Environment, ApJ, 681, 197-224.

24. Dunkley, J. et al. (2008). Five-year Wilkinson Microwave Anisotropy Probe (WMAP) Observations: Likelihoods and Parameters from the WMAP data, subm. ApJS; arXiv:0803.0586v1.

25. Bershadskii, A., and Sreenivasan, K.R. 2002. Multiscaling of cosmic microwave background radiation, Phys. Lett. A, 299, 149-152.

26. Bershadskii, A., and Sreenivasan, K.R. 2003. Extended self-similarity of the small-scale cosmic microwave background anisotropy Phys. Lett. A, 319, 21-23.

27. Bershadskii, A. 2006. Isotherms clustering in cosmic microwave background, Physics Letters A, 360, 210-216.

28. Tran, H. D., Sirianni, M., & 32 others 2003. Advanced Camera for Surveys Observations of Young Star Clusters in the Interacting Galaxy UGC 10214, ApJ, 585, 750.

29. Toomre, A., & Toomre, J. 1972. Galactic Bridges and Tails, ApJ, 178, 623.

30. Burbidge, G. 2003. NGC 6212, 3C 345, and other quasi-stellar objects associated with them, ApJ, 586, L119-L122.